\documentclass[12pt]{iopart}

\begin{document}

\title[Contributions of Vacuum and Plasmon Modes]
{Contributions of Vacuum and Plasmon Modes \\ to the Force on a Small
Sphere near a Plate}

\author{ L.H. Ford}

\address{Institute of Cosmology  \\
Department of Physics and Astronomy\\ 
         Tufts University, Medford, MA 02155}

\ead{ford@cosmos.phy.tufts.edu} 

\begin{abstract}
The force on a small sphere with a plasma model dielectric function and
in the presence of a perfectly reflecting plane is considered. The contribution
of both the vacuum modes of the quantized electromagnetic field and of plasmon
modes in the sphere are discussed. In the case that the plasmon modes are
in their ground state, quasi-oscillatory terms from the vacuum and plasmon
parts cancel one another, leading a monotonic attractive force. If the plasmon
modes are not in the ground state, the net force is quasi-oscillatory. 
In both cases, the sphere behaves in the same way as does an atom in either
its ground state or an excited state.
\end{abstract}

\section{Introduction}
\label{sec:intro}

As is well-known, Casimir and Polder~\cite{CP48} calculated the interaction
energy between an atom in its ground state and a perfectly reflecting plate
and obtained the result
\begin{equation}
V = - \frac{1}{4 \pi z^3}\, \int_0^\infty d\xi \, \alpha(i \xi) \,
 (2  z^2 \xi^2 + 2 z \xi +1)\, {\rm e}^{-2 z \xi} \, , \label{eq:CP}
\end{equation}
where $z$ is the distance between the atom and the plate and $\alpha(i \xi)$
is the atomic polarizability evaluated at imaginary frequency. In the limit
of large $z$, this potential takes the simple form
\begin{equation}
V \sim V_a = - \frac{3\, \alpha(0)}{8 \pi\, z^4}\,,   \label{eq:CPA}
\end{equation}
which can be interpreted as due to the interaction of the atom with
the quantized electromagnetic field. In particular, the asymptotic 
Casimir-Polder potential can be expressed as 
\begin{equation}
 V_a = - \frac{1}{2} \alpha(0) \, \langle {\mathbf E}^2 \rangle =
\frac{\alpha(0)}{4 \pi \, z^3}\, \int_0^\infty d \omega\,
 \Bigl[(2\,\omega^2\,z^2 -1) \sin 2\omega z
+ 2\,\omega\,z \cos 2\omega z \Bigr] \,.  \label{eq:Va}
\end{equation}
Here $\langle {\mathbf E}^2 \rangle$ is the shift in the mean square
of the quantized electric field due to the plate, and the integral on
$\omega$ is understood to be evaluated by use of a convergence factor.
For example, insert a factor of ${\rm e}^{- \beta \omega}$ in the integrand,
and then take the limit $\beta \rightarrow 0$ after evaluation of the integral.
The integrand is a function which is highly oscillatory, and the area under
each oscillation is much larger than the final result for the integral.
Thus, if one could slightly modify the frequency spectrum, which is the 
integrand in Eq.~(\ref{eq:Va}), both the magnitude and sign of the result
could be dramatically altered. The Casimir effect between parallel plates
also exhibits a quasi-oscillatory spectrum~\cite{Ford88,Hacyan}, and
the possible effects of its modification were discussed in 
Ref.~\cite{Ford93}.

In a previous paper, Ref.~\cite{F98}, the force on a small dielectric 
sphere in the presence of a perfectly reflecting plate was discussed.
It was shown that the interaction of the sphere with the quantized 
electromagnetic field produces a quasi-oscillatory force whose magnitude
can be much larger than the force associated with the asymptotic 
Casimir-Polder potential, Eq.~(\ref{eq:CPA}). This result suggests
a modification of the frequency spectrum of vacuum fluctuations.
More recently, the model in Ref.~\cite{F98} was extended to include
finite reflectivity of the wall and finite temperature effects~\cite{SF04}.
However, Barton~\cite{Barton04} has suggested that the results in 
Refs.~\cite{F98} and \cite{SF04} are incomplete in that they do not 
include a contribution from the quantum charge fluctuations 
in the sphere, and that this effect will cancel the quasi-oscillatory terms
when the sphere is in its ground state. 
 This contribution, also known as radiation reaction
or the effect of plasmonic modes, arises from the retarded self-interaction
of  quantum charge fluctuations in the presence of the mirror.  

As will be discussed below, Barton's suggestion is correct; the 
quasi-oscillatory terms cancel for the ground state, but not for other 
quantum states of the plasma in the sphere, which is essentially a three
dimensional quantum harmonic oscillator. In this way, the sphere behaves
just as does an atom. An atom in its ground state has the attractive
interaction found by Casimir and Polder, Eq.~(\ref{eq:CP}), but an atom
in an excited state has a quasi-oscillatory interaction potential with the
plate~\cite{Barton74}.

In Sect.~\ref{sec:vacuum}, the vacuum mode contribution found in 
Ref.~\cite{F98} will be reviewed. The contribution of the charge fluctuations
will be computed in Sect.~\ref{sec:plasmon}, and the results will be 
discussed in Sect.~\ref{sec:final}.

\section{Contribution of the Quantized Electromagnetic Field}
\label{sec:vacuum}

In this section, we will summarize the results of Ref.~\cite{F98}  
concerning the 
force due to the vacuum modes, or the quantized electromagnetic field.
If one has a polarizable particle with dynamic polarizability 
$\alpha(\omega)$
located a distance $z$ from a perfectly reflecting plane, the force on the
particle may be computed from the Maxwell stress tensor. The vacuum modes
in the presence of the plate consist of a linear combination of an incident and
a reflected wave. Each of these interacts with the particle and creates a
radiated dipole field. The force on the particle arises from cross terms
between (1) the  reflected wave and the dipole field induced by the incident
wave, and (2) the incident wave and the dipole field induced by the reflected
wave. After integrating over all modes, one finds the net force to be
\begin{eqnarray}
F_V &=&   - \frac{1}{4\pi z^4} \int_0^\infty d \omega \, \alpha_1(\omega)\,
   \nonumber \\   
 &\times&    \Bigl[ 3\sin 2\omega z -6 z\,\omega\, \cos 2\omega z
 -6 z^2\,\omega^2\,\sin 2\omega z
  + 4z^3\,\omega^3\,\cos 2\omega z \Bigr] \,.  \label{eq:force4}
\end{eqnarray} 
Here $\alpha_1(\omega)$ is the real part of $\alpha(\omega)$.
The appearance of the real part of the dynamic polarizability is a crucial 
feature of this result for $F_V$. Equation~(\ref{eq:force4}) is derived
using a dipole approximation which requires the sphere to be small compared
to both the distance to the plane, $z$, and to the characteristic wavelengths 
associated with $\alpha_1(\omega)$. 

Now consider a sphere of radius $a$ composed of a uniform material with
dielectric function $\varepsilon(\omega)$. The complex polarizability is given
by
\begin{equation}
\alpha(\omega) = a^3\, \frac{\varepsilon(\omega) - 1}{\varepsilon(\omega) +2}
                              \,.  \label{eq:alpha}
\end{equation}
We will take the dielectric function to be that of the Drude model,
\begin{equation}
\varepsilon(\omega) = 1 - \frac{\omega_p^2}{\omega(\omega + i\gamma)} \,,
                                                \label{eq:epsilon}
\end{equation}
where $\omega_p$ is the plasma frequency and $\gamma$ is the damping parameter.
From Eqs.~(\ref{eq:alpha}) and (\ref{eq:epsilon}), we find that the real part 
of the polarizability is given by
\begin{equation}
\alpha_1 =  a^3\, \omega_p^2\, 
\frac{\omega_p^2 -3 \omega^2}{(3 \omega^2 -\omega_p^2)^2 + 9 \omega^2 \gamma^2}
                               \,.  \label{eq:alpha_1}
\end{equation}
Note that although $\alpha(\omega)$ has poles only in the lower half-$\omega$
plane, its real part $\alpha_1(\omega)$ has poles in both the upper and lower
half planes. In the context of this model, the sphere is required to be small
in the sense that $a \ll z$ and 
\begin{equation}
a\, \omega_p \ll 1 \,.
\end{equation}

We can evaluate the integral in Eq.~(\ref{eq:force4}) 
by rotating the contour of integration
to the imaginary $\omega$-axis.
However, in this process we will also acquire
a contribution from the residue of the pole of $\alpha_1(\omega)$ at
$\omega = \Omega + \frac{1}{2}\, i\, \gamma$, where 
 \begin{equation}
\Omega = \frac{1}{6} \sqrt{12 \omega_p^2 - 9 \gamma^2} \,.
\end{equation}
Here we are primarily interested in the limit of small damping, so we 
will take the limit $\gamma \rightarrow 0$ after the integrals are evaluated.
In this limit,
 \begin{equation}
\Omega = \frac{\omega_p}{\sqrt{3}} \,. \label{eq:Omega}
\end{equation}

The force can be written as
\begin{equation}
F_V = J_V + P_V \,,
\end{equation}
where $J_V$ is the net contribution from integrals along the imaginary axis, 
and $P_V$ is that from the pole.  
The explicit forms of these two contributions for $\gamma = 0$ are
\begin{equation}
J_V = - \frac{a^3\, \omega_p^2}{4 \pi \, z^4}\, \int_0^\infty d \xi \,
\frac{(3 \xi^2 +\omega_p^2)(4 z^3 \xi^3 +6z^2 \xi^2 +6z\xi +3)}
     {(3 \xi^2 +\omega_p^2)^2 }\;  e^{-2 z \xi} \,,
                                                        \label{eq:J}
\end{equation}
and
\begin{equation}
P_V = - \frac{a^3\, \omega_p^2}{24 \,\Omega\, z^4}\, 
\Bigl[ 2\Omega \,z\,(2\,\Omega^2\, z^2 -3) \sin (2\Omega z) 
 + 3\,( 2\Omega^2\, z^2  -1) \cos (2\Omega z) \Bigr] \,.  \label{eq:P}
\end{equation}
Here  the positive direction is taken to be away from the plate, so
a positive value for the force indicates repulsion.

\section{Contribution of the Quantum Charge Fluctuations}
\label{sec:plasmon}

In this section, we will calculate the average force on the sphere
due to quantum fluctuations of the charge density in the sphere.
Here we consider only the case where $\gamma = 0$, so the plasma
oscillations are undamped, and have a resonate frequency at the pole of
$\epsilon(\omega)$, that is at $\omega = \Omega$ where $\Omega$ is given by
Eq.~(\ref{eq:Omega}).
Classically, the plasma forms a three-dimensional harmonic oscillator
with a frequency $\Omega$. When the electron gas is described quantum 
mechanically, the plasma becomes a  three-dimensional quantum harmonic 
oscillator. 

First, let us consider  a classical oscillating dipole in the presence 
of a perfectly reflecting plate located at $z = 0$. Let the instantaneous 
dipole moment be given by
\begin{equation}
\mathbf{p}(t) = \mathbf{p_0}\, \cos \Omega t \,.
\end{equation}
The electromagnetic field radiated by this dipole is reflected by the plate
and returns to exert a reaction force on the dipole. Let $\mathbf{E}(t)$
and $\mathbf{B}(t)$ be the reflected electric and magnetic fields  at the
location of the dipole. The resulting force on the dipole is given by
\begin{equation}
\mathbf{F}_P = p^j\, \partial_j\, \mathbf{E} + \mathbf{\dot{p} \times B}\,.
\end{equation}
(This form can be derived directly from the Lorentz force law, and differs
slightly from the expression for the force used in Ref.~\cite{F98}, 
which was derived from the Maxwell stress tensor. However, one can show that
both expressions lead to the same time-averaged force.)

We are interested in the component of time-averaged force which is normal
to the plate. A lengthy but straightforward calculation leads to the 
result
\begin{eqnarray}
F_P &=& \frac{p_{0x}^2 + p_{0y}^2}{32 \, z^4} \, 
\Bigl[ (8 \Omega^2 z^2 - 3)\, \cos(2 \Omega z)
+ 2 \Omega z\,(4 \Omega^2 z^2 -3 ) \sin (2 \Omega z) \Bigr]   \nonumber \\ 
 &+& \frac{p_{0z}^2}{16 \, z^4} \Bigl[(4 \Omega^2 z^2 -3) \, \cos(2 \Omega z)
-6 \, \Omega z \, \sin (2 \Omega z) \Bigr] \,.
\end{eqnarray}

Now we need to briefly summarize the quantum mechanical description
of the plasma in the sphere. Suppose that there are $N$ free electrons
in the sphere, and that each electron is displaced by $\mathbf{u}$
from its equilibrium position during the plasma oscillations. The
Lagrangian for the plasma may be written as
\begin{equation}
{\cal L} = \frac{1}{2} N m (\mathbf{\dot{u}}^2 - \Omega^2 \mathbf{u}^2) \,,
\end{equation}
where $m$ is the electron mass. Thus the system is equivalent to a single
particle with mass $N m$ is a three-dimensional harmonic potential  with
resonant frequency $\omega$. The dipole moment of the system can be expressed 
as
\begin{equation}
\mathbf{p} = N e \, \mathbf{u} \,,
\end{equation}
where $e$ is the electron charge.
When the system is  quantized,  the Cartesian components of  $\mathbf{u}$
can be expressed as
\begin{equation}
u_j = \frac{1}{\sqrt{2 N m \Omega}} \, (a_j + a^\dagger_j) \, ,
\end{equation}
where $a_j$ and $a^\dagger_j$ are the annihilation and creation operators,
respectively, for quanta in mode $j$. 

We can make the link between the previous description of a classical
oscillating dipole and a dipole undergoing quantum fluctuations by
identifying  $\langle p_j^2 \rangle$, 
the quantum expectation value of $p_j^2$, with the time-average
of the corresponding squared classical field. Thus, we can set
\begin{equation}
p_{0j}^2 = 2 \langle p_j^2 \rangle \,. 
\end{equation}

If the oscillator is in its ground state, we have
\begin{equation}
\langle p_x^2 \rangle_0 = \langle p_y^2 \rangle_0 = \langle p_z^2 \rangle_0
= \frac{1}{2 N M \Omega} = \frac{a^3 \, \omega_p^2}{6 \Omega} \,,
\end{equation}
where we used the fact that
\begin{equation}
\omega_p^2 = \frac{4 \pi N e^2}{V m} = \frac{3 N e^2}{ a^3 m} \, ,
\end{equation}
where $V$ is the volume of the sphere.
Thus in the ground state, we find that $F_P = -P_V$. The force due to
the averaged charge fluctuations exactly cancels the quasi-oscillatory 
term coming from the vacuum modes. The net force is then
\begin{equation}
F = J_V +P_V  + F_P = J_V \,,
\end{equation}
which is always attractive. 

The situation is quite different of the quantum oscillator is not in its 
ground state. In this case, we can write the net force as
\begin{eqnarray}
F &=& J_V + \frac{1}{16 \, z^4} \,
\Bigl(\langle p_x^2 \rangle - \langle p_x^2 \rangle_0 +
\langle p_y^2 \rangle - \langle p_y^2 \rangle_0 \Bigr)
\Bigl[ (8 \Omega^2 z^2 - 3)\, \cos(2 \Omega z)  \nonumber \\
&+& 2 \Omega z\,(4 \Omega^2 z^2 -3 ) \sin (2 \Omega z) \Bigr]   \nonumber \\ 
 &+& \frac{1}{8\, z^4} \,
\Bigl(\langle p_z^2 \rangle-\langle p_z^2 \rangle_0\Bigr)
\Bigl[(4 \Omega^2 z^2 -3) \, \cos(2 \Omega z)
-6 \, \Omega z \, \sin (2 \Omega z) \Bigr] \,. \label{eq:genstate}
\end{eqnarray}
Consider, for example, the case where the state of the oscillator contains
one quantum in a particular mode. For that mode, we will have
$\langle p_j^2 \rangle = 2 \langle p_j^2 \rangle_0$, and $F$ will have a 
quasi-oscillatory part which is just minus the vacuum contribution.
Note that Eq.~(\ref{eq:genstate}) is equivalent to 
Barton's~\cite{Barton74} Eq.~(4.1).

A case of special interest is when the oscillator is in a coherent state
of each mode. In this case,
we have
\begin{equation}
a_j |\psi \rangle = \beta_j |\psi \rangle \,,
\end{equation}
 where $|\psi \rangle$ is the quantum state of the oscillator, and the
$\beta_j$ are three complex numbers. In this case, the net force becomes
\begin{eqnarray}
F &=& J_V + \frac{2 \sqrt{3} a^3 \omega_p}{48 \, z^4} \,
\biggl\{\Bigl[ ({\rm Re} \beta_x)^2 +  ({\rm Re} \beta_y)^2 
\Bigl[ (8 \Omega^2 z^2 - 3)\, \cos(2 \Omega z)  \nonumber \\
&+& 2 \Omega z\,(4 \Omega^2 z^2 -3 ) \sin (2 \Omega z) \Bigr]   \nonumber \\ 
 &+&  2\, ({\rm Re} \beta_z )^2
\Bigl[(4 \Omega^2 z^2 -3) \, \cos(2 \Omega z)
-6 \, \Omega z \, \sin (2 \Omega z) \Bigr] \biggr\}\,.
\end{eqnarray}
Thus if the quantum state of the plasma is a coherent state with $|\beta_j|$
of order one, then there is a quasi-oscillatory force on the sphere which
is of the same order of magnitude as the vacuum contribution, $F_V$.

\section{Discussion and Conclusions}  
\label{sec:final}  

We have seen that plasmonic contributions, due to quantum charge fluctuations,
give a significant  contribution to the force on the sphere. When these modes 
are in their ground state, they exactly cancel the quasi-oscillatory terms
$P_V$ arising from the vacuum modes. In this case, only the vacuum mode
term $J_V$ which can be written as an integral over imaginary frequency
remains. Thus if one had ignored both the plasmon mode contribution and
the residues of the poles of $\alpha_1(\omega)$, one would have inadvertently 
obtained the correct answer. When the plasmon modes are not in the ground 
state, the net force is quasi-oscillatory. 

In the context of atomic systems, there is an extensive literature on
the relative effects of vacuum modes and quantum charge fluctuations,
often called radiation reaction or self-reaction. See, for example,
Ref.~\cite{Dalibard} and \cite{MJH90}. However, many of the treatments
for atomic systems do not make it clear whether the separation between vacuum
and self-reaction effects is gauge invariant. In the present model, it
is obvious that the two effects are separately gauge invariant, as all
of the calculations involve field strengths rather than potentials.
When calculations are performed by doing second order perturbation theory
on a Hamiltonian, as in Refs.~\cite{CP48} and \cite{Barton74}, there is
no need to consider the effects separately, and the cancellation is 
automatic.

In the case of the Casimir effect between planar geometries, the role
of plasmonic contributions at short distances has long been 
recognized~\cite{VanKampen,Barton79}.
More recently, Intravaia and  Lambrecht~\cite{IL05} have argued that they 
are significant at all separations. 

It seems to be desirable to understand better the physical reason for
the cancellation between vacuum and plasmonic effects.  It is clear that 
one way to upset the cancellation is to excite the plasmon modes.
However, it is less clear if this is the only way in more general
geometries. 

\ack
I would like to thank Gabriel Barton for valuable criticism and discussions.
This work was supported in part by the National Science Foundation under
Grant PHY-0244898.

\end{document}